\documentclass{article}
\usepackage{amssymb}
\usepackage{amsmath}

\begin{document}
\date{}

\title{GSM Security Using Identity-based Cryptography}
\author{Animesh Agarwal, Vaibhav Shrimali, and Manik Lal Das\\
Dhirubahi Ambani Institute of\\ Information and Communication Technology\\
Gandhinagar - 382007, India.\\\{animesh\_agarwal, vaibhav\_shrimali,
maniklal\_das\}@daiict.ac.in}

\date{}
\maketitle

\begin{abstract}
Current security model in Global System for Mobile Communications
(GSM) predominantly use symmetric key cryptography. The rapid
advancement of Internet technology facilitates online trading,
banking, downloading, emailing using resource-constrained handheld
devices such as personal digital assistants and cell phones.
However, these applications require more security than the present
GSM supports. Consequently, a careful design of GSM security using
both symmetric and asymmetric key cryptography would make GSM
security more adaptable in security intensive applications. This
paper presents a secure and efficient protocol for GSM security
using identity based cryptography. The salient features of the
proposed protocol are (i) authenticated key exchange; (ii) mutual
authentication amongst communicating entities; and (iii) user
anonymity. The security analysis of the protocol shows its
strength against some known threats observed in conventional GSM
security.\vspace{2 mm}
\end{abstract}

\section{Introduction}
Global System for Mobile Communication (GSM) \cite{hal03} has been
the most widely used technology in connecting mobile devices
(e.g., Personal Digital Assistants (PDAs), cell phones)
wirelessly, where mobile devices remain connected even on the move
and irrespective of their geographical locations. The GSM
technology was introduced as an improvement over analog First
Generation wireless systems. Earlier GSM was used to transfer only
voice and text conversations, but with the introduction of
services like mobile banking, secure billing, we surfing, so on
and so forth, there is a need to update the current GSM security
model. As a result, additional security attributes, such as mutual
authentication, dynamic key establishment, non-repudiation are
important requirements for securing these applications. However,
to the best of authors' knowledge, the main challenges in
designing GSM security protocol are i) providing higher level of
security while keeping the computational and communication cost
low on resource-constrained mobile devices, and ii) compatibility
issues amongst devices and applications.\\
The security philosophy of GSM was driven by the concerns to
ensure the following: \cite{bro}.
\newcounter{xx}
\begin{list}{-}
{\usecounter{xx}} \item Anonymity of subscriber: Subscriber's
identity, which is a unique number IMSI, is not sent through
over-the-air channel in clear, for this a temporary identity TMSI
is used through which network retrieves IMSI. IMSI is always sent
through a secure channel. \item Authentication of subscriber:
authentication of the subscriber is done through challenge
response mechanism. \item Confidentiality of user and signaling
data: in the current GSM scheme a secret key between user and
network is established to communicate through over-the-air
insecure channel which ensures confidentiality of user and
signaling data. \item One-way authentication: Current GSM security
model does not provide mutual authentication. \item The channel
between VLR and HLR is assumed to be a secure private channel.
\end{list}
Having observed the merits and limitations of the GSM network, we
think the GSM security model has to be evolved in such a way that
the network should be used not only in voice communications or
text messaging but also it should support adequate security
strength for accommodating secure billing, trading, content
up/downloading, etc. This paper proposes a scheme which aims at
achieving the same. It uses ID-based cryptosystem and Elliptic
Curve Cryptography, these two techniques allow us to achieve high
level of security at reduced key lengths. In the proposed scheme
the resource constraint mobile device does not indulge in any
computationally extensive task, thereby providing a high level of
performance.

\subsection {GSM Architecture}
There are three main components of GSM architecture:\\
\newcounter{11}
\begin{list}{-}
{\usecounter{11}} \item The Mobile Station (MS): It consists of
mobile wireless equipment (i.e. the hardware) and the subscriber
information. The subscriber information is the IMSI (International
Mobile Subscriber Identity) which is stored in the SIM (Subscriber
Identity Module). \item Base Station Subsystem (BSS): It consists
of Base Transceiver Station (BTS) and Base Station Controller
(BSC). The BTS contains radio transceivers and engages in radio
link protocols with MS, the BSC controls and manages radio
resources of several BTSs. Also BSC is responsible for radio
channel setup, frequency hopping and handovers between two BTSs
that the BSC controls. \item Network Subsystem (NSS): It consists
of five entities:
\begin{itemize}
\item Mobile Switching Centre (MSC): It provides functionalities
like registration, authentication, location updating and call
routing to a roaming subscriber, to the MS. \item Home Location
Register (HLR): It contains the current location of the subscriber
registered in the GSM network and manages administrative
information about the MS. \item Visitor Location Register (VLR):
The VLR is an agent deputed by HLR with some administrative
responsibilities for services to the MS. \item Equipment Identity
Register (EIR): It is a database that stores the list identities
of all valid mobile stations in the network. \item Authentication
Centre (AuC): It is a protected database containing the shared
secret keys stored in the mobile subscriber's SIM.
\end{itemize}
\end{list}

\subsection{Paper contributions}
The current GSM security model is based on the symmetric key
cryptography, which does not support required security services
for security intensive applications such as mobile banking,
trading, emailing, etc. The paper is aimed at securing GSM network
using identity-based cryptography. The proposed protocol does not
require any secure/dedicated channel between the BSS and network
servers. As the mobile devices are resource constraint and have
limited computing capability, the protocol has been designed in
such a manner that mobile devices do not require to compute costly
public key operation. The protocol provides mutual authentication,
anonymity, replay protection, and mutual key control for
establishing session key. After successful authentication between
mobile device and authentication server, the transmitted data are
being encrypted under a dynamic transient key established by the
MS and VLR/HLR.

\subsection{Paper organization}
In section II, we discuss preliminaries that would make the paper
self sufficient, followed by related work. In section III, we
present our protocol. In section IV, we analyze the proposed
protocol. We conclude the paper with Section V.

\section {Preliminaries and Related Work}
In this section we discuss some preliminary concepts that are
required for thorough understanding of the proposed protocol. It
includes the basic concepts of Elliptic Curve Cryptography and
Bilinear Pairing. After that we would discuss the related work
done in the fields of `Authentication in GSM' as well as
'ID-based cryptosystems'.\vspace{2 mm}\\
\textbf{Elliptic curve.} An elliptic curve $E$ over a field $F$
(in short, $E(F)$) is a cubic curve \cite{han03} with no repeated
roots. The general form of an elliptic curve is
$Y^2+a_1XY+a_3Y=X^3+a_2X^2+a_4X+a_5$, where $a_i \in F$,
$i=1,2,\cdots,5$. The $E(F)$ contains the set of all points
$P(x,y)$ on the curve, such that $x$, $y$ are elements of $F$
along with an additional point called \emph{point at infinity}
($\mathcal{O}$). The set $E(F)$ forms an Abelian group under
elliptic curve point addition operation with $\mathcal{O}$ as the
additive identity. The addition rules are as follows: For all $P,Q
\in E(F)$, let $F_q$ be a finite field with order $q$ and
characteristic $p$.\vspace{ 2mm}\\
\textbf{Point Addition.}
\begin{enumerate}
\item If $P = (x_1, y_1) \neq \mathcal{O}$, then $-P =
(x_1,-y_1)$. \item If $ P \neq Q$ and $ P,Q \neq \mathcal{O}$,
then the line joining $P$ and $Q$ intersects the curve in another
point $R$. \item If $P = Q$ and $P, Q \neq \mathcal{O}$, then the
line $l$ is the tangent at $P$ and intersects the curve at another
point $-R$. Then define $P+Q = 2P = R$.
\end{enumerate}
\textbf{Scalar Multiplication of a Point.} The scalar, $n$,
multiplication of a curve point $P$ is defined as $n$-times
addition of $P$, i.e., $nP = P + P + \cdots + P$ ($n$-times).
There are algorithms \cite{dim05,kob94} for faster computation of
scalar multiplication of a curve point.

\subsection{Computational Problems} Let there is a
randomized parameter generator algorithm and a polynomial time
algorithm that takes as input a security parameter 1$^k$ and
outputs the required/resultant parameters.\vspace{2 mm}\\
\textbf{Discrete Logarithm Problem (DLP).} Let $p$ and $q$ be two
prime numbers such that $q|(p-1)$. Let $g$ be a random element
with order $q$ in $Z_p^*$, and $y$ be a random element generated
by $g$. Then, for any probabilistic polynomial time algorithm
$\mathcal{P}$, the probability that $\mathcal{P}$($p, q,
g, y) = x$ such that $g^x = y$ mod $p$ is a negligible function in $k$.\vspace{2mm}\\
\textbf{Elliptic Curve Discrete Logarithm Problem (ECDLP).} Given
an elliptic curve $E(F_q)$, points $P$ and $Q$(=$xP$) for $P, Q
\in E(F_q)$, the ECDLP is to determine the integer $x$.\vspace{2mm}\\
\textbf{Computational Diffie-Hellman Problem (CDHP).} Given $(P,
aP, bP)$ for all $a, b \in \mathbb Z_q^*$, compute $abP$. The
advantage of any probabilistic polynomial-time algorithm
$\mathcal{P}$ in solving CDHP in $G_1$, is defined as
\texttt{Adv}$_{{\mathcal{P}}, G_1}^{CDH}$ =
\texttt{Pr}[${\mathcal{P}}(P, aP, bP, abP)=1$ for all $a,b \in
\mathbb Z_q^*$]. And the advantage is negligible in $k$ for all
probabilistic polynomial time algorithms.

\subsection{Bilinear Pairings}
A bilinear pairing \cite{bon01} is a function that takes as input
two group elements and outputs an element of a multiplicative
Abelian group. The Weil pairing and the Tate pairing are the two
most commonly used bilinear pairings in cryptography. So far, a
large number of schemes and protocols have been proposed using
bilinear pairings, and we refer to \cite{par} for various pairing
based cryptographic schemes and protocols. Formally, the bilinear
pairing is defined as follows:\\
Suppose $G_1$ is a cyclic additive group of prime order $q$, $G_2$
is a cyclic multiplicative group of the same order and $P$ be a
generator of $G_1$. A map $\hat{e} : G_1 \times G_1 \rightarrow
G_2$ is called a bilinear pairing if it satisfies the following
properties:\\
- Bilinearity: $\hat{e}(aP, bQ) = \hat{e}(P,Q)^{ab}$ for all $P,Q
\in G_1$ and $a, b \in \mathbb Z_q^*$.\\
- Non-degeneracy: There exist $P, Q \in G_1$ such that
$\hat{e}(P,Q) \ne 1$.\\
- Computability:  There is an algorithm to
compute $\hat{e}(P,Q)$ for all $P, Q \in G_1$.\\
In general, $G_{1}$ is the group of points on an elliptic curve
and $G_{2}$ denotes a multiplicative subgroup of a finite field.
One may refer to \cite{par} for an informative description of
various parameters involved in bilinear pairings and its
implementation details.

\subsection{Map-to-Point}
Map-to-Point (a special hash function) is an algorithm for
converting an arbitrary bit string onto an elliptic curve point.
Firstly, the string has to be converted into an integer and then a
mapping is required from that integer onto an elliptic curve
point. As our scheme uses this special hash operation, we discuss
a basic algorithm for Map-to-Point \cite{bon01} as follows.\\
Let $E(F_p)$ be an elliptic curve defined as $y^2 = f(x)$ and let
$F_p$ has order $m$. Let $P \in E(F_p)$ be a point of prime order
$q$ and $h: \{0,1\}^* \rightarrow F_p \times \{0,1\}$ be a one-way
collision-resistant hash function. The algorithm works as follows:\\
1) Given $M \in \{0,1\}^*$, Set $i=1$\\
2) Set ($x, b$) = $h(i \| M) \in F_p \times \{0,1\}$\\
3) If $f(x)$ is a quadratic residue in $F_p$ then

-- Let $y_0, y_1 \in F_p$ be the two square roots of $f(x)$

-- Set $\hat{P_M} = (x, y_b)$ such that $y_b$ = \texttt{Max}$(y_0, y_1)$\\
4) Compute $P_M = (m/q)\hat{P_M}$. Then $P_M \in E(F_p)$.\\
5) Otherwise increment $i$ and go to step 2.

\subsection{GSM Security}
The GSM security \cite{bro,bri98} is based on A3, A5, and A8
algorithms using a master secret generated by the NSS and stored
at MS and network's authentication server. A5 is based on
combination of three 'Linear Feedback Shift Registers(LFSR)'.
These are initialized with a 64-bit key whose 10-bits are always
zero, therefore the effective key length is 54-bits.

\subsubsection{Registration Phase: }
During MS registration, AuC registers a 128 bit key Ki and stores
it in the SIM of the MS

\subsubsection{Authentication Phase}
\newcounter{a}
\begin{list}{-}
{\usecounter{a}}
\item MS $\rightarrow$ VLR :$<$TMSI$>$ \\
Whenever a MS want to connect to the GSM network it sends the
nearest VLR its TMSI. Since the subscriber is sending TMSI and not
IMSI, he remains anonymous.
\item VLR $\rightarrow$ HLR : $<$IMSI$>$ \\
If MS is not in roaming then VLR itself knows IMSI but if the MS
is on roaming scenario then VLR has to send the received TMSI to
the previous VLR under whom the MS was held.\\
Upon having IMSI, VLR sends the IMSI to HLR.
\item HLR $\rightarrow$ VLR: $<$RAND,SRES,$K_{c}$ $>$\\
HLR sends back five distinct authentication triplets to VLR. Each
triplet consists of three entities, namely, a random number
(RAND), a signed response (SRES =A3(RAND, $K_{i}$)), and an
encryption key ($K_c$ = A8(RAND, $K_{i}$)).
\item VLR $\rightarrow$ MS: $<$RAND$>$ \\
VLR selects one triplet to authenticate the MS and sends RAND to
the MS.
\item MS $\rightarrow$ VLR : $<$SRES'$>$ \\
MS computes SRES'= A3 (RAND,$K_{i}$) and $K'_{c}$  =
A8(RAND,$K_{i}$). MS then sends back SRES' to the VLR \item VLR
upon receiving SRES' compares it with SRES in the authentication
triplet. If SRES = SRES' then MS is authenticated else VLR
terminate the operation.
\end{list}


\subsection{Public key cryptography in GSM}
GSM security using public key cryptography started back in 1989
with Yacobi and Shimley \cite{yac89} proposed a key distribution
protocol which is based on Diffie-Hellman \cite{dif76}, but this
protocol is vulnerable to impersonation attack \cite{hor02}.
Bellar and Yacobi \cite{bel93}proposed a protocol based on ElGamal
encryption \cite{elg85}, but later this protocol was also proven
insecure \cite{par97}. Subsequently several protocols
\cite{boy98}, \cite{asp}, \cite{lee03}, \cite{gre03} have been
proposed securing mobile communications using public key
cryptography.

\section{The Proposed Protocol}
The proposed protocol uses the fascinating features of
identity-based cryptography, where the security of the protocol is
based on Computational Diffie-Hellman Problem (CDHP) cited in
section II. In 1984, Shamir \cite{sha84} proposed the first ID
based cryptosystems and signature scheme. ID based cryptosystem
has a property that the public key can be derived from an identity
that can help us identify any user uniquely thus this scheme
obviates
the need of any public key certificates.\\
The proposed protocol allows both MS and NSS to authenticate each
other, then they establish a transient secret key for securing
data to be transmitted in that session. One of the interesting
feature of the proposed protocol is that in every session, MS does
only hashing, XORing, and one scalar multiplication on elliptic
curve point,
whereas, HLR and VLR computes public key operations.\\
The protocol consists of three phases - setup, registration, and
authenticated key exchange. The setup and registration phases are
a one-time operation, but the authenticated key exchange phase is
a dynamic operation, as and when demanded. The symbols and
notation used in our protocol are given in Table 1.\\

\begin{table}[ht!]
\centering
\begin{tabular} {|r||l|}
\hline Notions & Explanations \\\hline $IMSI$ & International
Mobile Subscriber Identity\\
$TMSI$ & Temporary Mobile Subscriber Identity\\
$VLR_{ID}$ & $VLR$'s identity\\
$HLR_{ID}$  & HLR's identity\\
$HLR_{pub}$ & $HLR$'s public Key \\
$HLR_{pri}$ & $HLR$'s private Key \\
$H(\cdot)$ & Map-to-point, a special hash function.\\
MS$_{pub}$ & MS's public key \\
A $\rightarrow$ B : $m$ & message $m$ is sent from an entity A to B \\
$m_1 \oplus m_2$ & Bit-wise XORing of strings $m_1$ and $m_2$ \\
$m_1 \| m_2$ & Concatenation of strings $m_1$ and $m_2$\\
$E_{X}(m)$& message $m$ is encrypted using the key $X$\\
$SIGN_{X}(m)$& message $m$ is signed using the key $X$\\
\hline
\end{tabular}
\caption{Symbols and Notation}
\end{table}

\subsection{Setup Phase.} HLR chooses a random
integer $s \in$ [1, p-1], where $p$ is large prime number. Then
HLR selects a master secret key $K$ and generates a public-private
key pair ($HLR_{pub}, HLR_{pri}$) as $HLR_{pri} = K \cdot
H(HLR_{ID}$) and $HLR_{pub} = H(HLR_{ID}$), where $HLR_{pri}$ and
$K$ are secret, and $HLR_{pub}$ and $H(\cdot)$ are made public.

\subsection{Registration Phase.} When a new MS wants to register
with NSS, the HLR of that NSS generates a key $K' = K \cdot
H(IMSI)$ corresponding to $IMSI$ of the MS. It, then, stores this
$K'$ in the $SIM$ of the MS and also stores a copy of $K'$ with
itself for further verification of the MS.

\subsection{Authenticated Key Exchange Phase.}
This phase enables both MS and VLR to authenticate each other and
after successful authentication they establish a transient session
key. All the following message transmissions are done over the
public channel. The phase works as follows:

\newcounter{b}
\begin{list}{\arabic{b})}
{\usecounter{b}}
\item MS $\rightarrow$ VLR : $<TMSI>$\\
When MS wants to use services of NSS it sends its $TMSI$. \item
VLR $\rightarrow$ MS
: $<RAND>$\\
Upon receiving MS's $TMSI$, VLR sends $RAND$ back to the MS, where
$RAND$ is a nonce used to protect the protocol against replay
attacks. \item MS $\rightarrow$ VLR : $H(K'') \| TMSI \|
RAND''$.\\
After receiving $RAND$, MS generates another nonce $RAND'$ and
computes $RAND''$ as $RAND'' = RAND \oplus RAND'$. Then, the MS
calculates $K'' = K' \cdot H(RAND'')$ and sends $H(K'') \| TMSI \|
RAND''$ to VLR. \item VLR $\rightarrow$ HLR : $E_{HLR_{pub}}
 (IMSI \| H(K'') \| VLR_{ID} \| RAND'')$.\\
VLR, now, obtains $IMSI$ for corresponding $TMSI$ it received from
MS. Then VLR creates a message $< IMSI \| H(K'') \| VLR_{ID} \|
RAND >$ and encrypts it using $HLR_{pub}$. Here, the encryption
algorithm could use any public key encryption algorithm. However,
as we intend to use identity based cryptography, the
identity-based encryption \cite{bon01} by Boneh and Franklin finds
an application in our protocol. \item HLR decrypts the message
using its private key and gets $K'$ corresponding to the $IMSI$ it
received. It then generates $K'' = K' \cdot H(RAND'')$ and checks
if calculated $H(K'')$ ia same as the received value. If it does,
it authenticates MS. It then authenticates VLR on seeing the
$VLR_{ID}$ it received. If either of VLR or MS is not
authenticated, the session is terminated. \item HLR $\rightarrow$
VLR : $SIGN_{HLR_{pri}}\{H(IMSI \| K'' \| VLR_{ID})\}$, $H( IMSI
\| K'' \| VLR_{ID}$).\\ HLR sends $SIGN_{HLR_{pri}}\{H(IMSI \| K''
\| VLR_{ID})\}$, $H(IMSI \| K'' \| VLR_{ID}$) to the VLR. VLR
first verifies the signature, and if the signature is valid then
VLR proceeds to next step; otherwise, terminate the session. Here,
the short-signature scheme \cite{bon02} by Boneh et al. is
applicable for signing the message.
\item VLR $\rightarrow$ MS: $H(IMSI \| K'' \| VLR_{ID}), VLR_{ID}$).\\
VLR appends its identity $VLR_{ID}$ to the message it received
from HLR and relays it to MS. \item MS generates $H(IMSI \| K'' \|
VLR_{ID}$). If the computed hashed matches the value which it
received from the VLR, MS authenticates the NSS (both HLR and
VLR); otherwise, the session is terminated.
\end{list}

\section{Security Analysis}
The proposed protocol has two communication channels - one in
between MS and VLR and other in between VLR and HLR. It is
important to note that both channels are public channels, that is,
all transmitted messages are available to the adversary. However,
we show that the protocol resists the following attacks, which are
potential threats of any authenticated key exchange protocol.

\subsection{Replay Attack}
Replay attack is an offensive attack in which an adversary
intercepts a session and then replays some of the intercepted
parameters to gain control over a new session. The proposed
protocol resists replay attacks, as two nonces are involved in
each session. The MS verifies the freshness of $RAND$ by seeing
its current value and previous value. Typically, if the current
value of $RAND$ is greater than the previous one, then MS proceeds
further, else, terminate the communication. It is also important
to note that one of the nonces is never sent over the network. As
a result, if adversary replays the same value of $RAND$, then
$K''$ will be different for different sessions. Therefore, replay
attack is not succeeded in our protocol.

\subsection{Mutual Authentication}
The protocol provides strong authentication. The communicating
entities (MS, VLR, HLR) involved in a session authenticate each
other before they agree on a shared session key. The
authentication process works as follows.
\newcounter{c}
\begin{list}{-}
{\usecounter{c}} \item \textbf{MS authenticating VLR and HLR.} MS
sends $H(K'')$ and $TMSI$ to VLR. It is the responsibility of the
VLR to obtain correct $IMSI$ for the corresponding $TMSI$ and of
the HLR to obtain the value of $K'$ for the received $IMSI$. These
two responsibilities cannot be done by any other parties other
than authentic VLR and HLR. Now, when the MS gets $H(IMSI \| K''
\| VLR_{ID})$ from the VLR (which came from the HLR) if the
verification of this hash holds correct then MS can be sure that
the $K''$ was correctly computed by the HLR. Consequently, both
VLR and HLR authenticity are confirmed. \item \textbf{HLR
authenticating MS and VLR.} HLR when receives\\
$E_{HLR_{pub}}\{IMSI \| H(K'') \| VLR_{ID} \| RAND''\}$ then HLR
authenticates the VLR by decrypting it and then checking the value
of $VLR_{ID}$ and also by verifying the value of $IMSI$. Further,
HLR authenticates MS by recomputing $K''$ and verifying the
$H(K'')$ it received. \item \textbf{VLR authenticating MS and
HLR.} The authenticity of HLR is achieved by its signature on
$H(IMSI \| K'' \| VLR_{ID})$ and MS's authenticity through its
$TMSI$, which should have a mapping to a valid $IMSI$ stored at
VLR's database.
\end{list}

\subsection{Anonymity}
The proposed protocol provides anonymity for the mobile entity
(MS). The MS can be identified on the network by the means of a
valid $IMSI$. The $IMSI$ of a MS is not transmitted in public,
instead, MS sends its $TMSI$ over the public channel. Upon
receiving $TMSI$, the VLR maps the $TMSI$ to a valid $IMSI$. If a
valid $IMSI$ corresponding to $TMSI$ is found in VLR's database
then VLR sends the $IMSI$ along with other parameter to HLR in an
encrypted manner using HLR's public key, so that HLR decrypts it
using his private key and get the MS's $IMSI$. As a result, MS's
identity is not disclosed any other party except VLR and HLR.


\subsection{Impersonation Attacks}
An adversary cannot impersonate MS and HLR for the following
reasons:\\
- MS computes $H(K'') \| TMSI \| RAND$ that requires to compute
$K''$, which in turn requires the secret key $K'$ known to MS.\\
- VLR and HLR secrecy is based on public key encryption.

\subsection{Mutual Key Control}
For each session, the protocol generates a new session key $K''$
which is computed as\\
$K'' = RAND'' \cdot K'$, where $RAND'' = RAND \oplus RAND'$.
Here $K'$ is generated by the HLR and is stored in the MS.\\
$RAND$ is generated by the VLR, and $RAND'$ is generated by the MS.\\
Therefore, in computation of a session key, the protocol requires
components from all the three participating entities. In other
words, no one is allowed to be biased in a particular session key,
and thus, the protocol provides mutual key control.

\subsection{Freshness of Session Key}
The protocol run resists the freshness property of the session if
the intruder cannot guess a fresh session key with a
non-negligible probability. In other words, at the end of the
protocol run the intruder should not be able to distinguish a
fresh session key, say $sk$, from a randomly chosen key from a
pool of session keys issued in some of the previous sessions. In
order to illustrate the protocol strength against freshness of the
session key, we consider the following challenger-intruder game,
proposed in \cite{bel94}.\\
Intruder general query: The intruder makes query to our protocol,
for some of the previously issued
session keys and obtains a set $\mathcal{S}$ of session keys.\\
Intruder special query: The intruder makes a special query to our
protocol for a fresh session key and obtains
$sk_{fh}$ as the fresh session key.\\
Challenger challenge: The challenger asks the intruder to answer
the freshness of a session key $sk_{ch}$ by running the protocol
or choosing a sample from $\mathcal{S}$. In other words, the query
is generated in flipping a fair coin $b \in \{0,1\}$ and returning
$sk_{ch}$ if $b = 0$, or else a random
sample from $\mathcal{S}$ if $b = 1$.\\
Intruder guess: The intruder now guesses $b$. Let intruder's guess
is $result \in \{0, 1\}$. The advantage that the intruder
correctly identifies whether he was given the fresh session key or
just a sample from $\mathcal{S}$ is max\{0, Pr[$result$] -
$\frac{1}{2}$\}. The protocol run resists the freshness property
of the session key if the intruder cannot guess session key's
freshness with a non-negligible probability.

\section{Conclusion}
We proposed a secure and efficient protocol for GSM security using
identity based cryptography. The proposed protocol does not
require any secure/dedicated channel between the BSS and network
servers. The protocol does not consume much computational resource
on mobile device and provides mutual authentication, anonymity,
replay protection, and mutual key control for establishing session
key. After successful authentication between mobile device and
authentication server, the communicating principals establish a
session key by which transmitted data is being protected. The work
could be extended in making the protocol secure against
denial-of-service and providing forward secrecy.

\bibliographystyle{latex8}

\end{document}